\documentclass[sigconf]{acmart}

\AtBeginDocument{%
  \providecommand\BibTeX{{%
    \normalfont B\kern-0.5em{\scshape i\kern-0.25em b}\kern-0.8em\TeX}}}

\copyrightyear{2022}
\acmYear{2022}
\setcopyright{acmlicensed}
\acmConference[AIES '22] {Proceedings of the 2022 AAAI/ACM Conference on AI, Ethics, and Society}{August 1--3, 2022}{Oxford, United Kingdom.}
\acmBooktitle{Proceedings of the 2022 AAAI/ACM Conference on AI, Ethics, and Society (AIES '22), August 1--3, 2022, Oxford, United Kingdom}
\acmPrice{15.00}
\acmISBN{978-1-4503-9247-1/22/08}
\acmDOI{10.1145/3514094.3534145}

\begin{document}
\fancyhead{}

%%
%% The "title" command has an optional parameter,
%% allowing the author to define a "short title" to be used in page headers.
\title[Crowdsourcing Impacts]{Crowdsourcing Impacts: Exploring the Utility of Crowds for Anticipating Societal Impacts of Algorithmic Decision Making}

%%
%% The "author" command and its associated commands are used to define
%% the authors and their affiliations.
%% Of note is the shared affiliation of the first two authors, and the
%% "authornote" and "authornotemark" commands
%% used to denote shared contribution to the research.
\author{Julia Barnett}
\email{juliabarnett@u.northwestern.edu}
\orcid{0000-0002-3476-1110}
\affiliation{%
 \institution{Northwestern University}
 \streetaddress{2240 Campus Drive, Room 2-168, Attention: Dept of Comm Studies
}
 \city{Evanston}
 \state{Illinois}
 \country{USA}
 \postcode{60208}
}

\author{Nicholas Diakopoulos}
\email{nad@northwestern.edu}
\orcid{0000-0001-5005-6123}
\affiliation{%
 \institution{Northwestern University}
 \city{Evanston}
 \state{Illinois}
 \country{USA}
 \postcode{60208}
}

\author{%Authors
        % All authors must be in the same font size and format.)
}

%%
%% By default, the full list of authors will be used in the page
%% headers. Often, this list is too long, and will overlap
%% other information printed in the page headers. This command allows
%% the author to define a more concise list
%% of authors' names for this purpose.
\renewcommand{\shortauthors}{Barnett and Diakopoulos}

%%
%% The abstract is a short summary of the work to be presented in the
%% article.
\begin{abstract}
  With the increasing pervasiveness of algorithms across industry and government, a growing body of work has grappled with how to understand their societal impact and ethical implications. Various methods have been used at different stages of algorithm development to encourage researchers and designers to consider the potential societal impact of their research. An understudied yet promising area in this realm is using participatory foresight to anticipate these different societal impacts. We employ crowdsourcing as a means of participatory foresight to uncover four different types of impact areas based on a set of governmental algorithmic decision making tools: (1) perceived valence, (2) societal domains, (3) specific abstract impact types, and (4) ethical algorithm concerns. Our findings suggest that this method is effective at leveraging the cognitive diversity of the crowd to uncover a range of issues. We further analyze the complexities within the interaction of the impact areas identified to demonstrate how crowdsourcing can illuminate patterns around the connections between impacts. Ultimately this work establishes crowdsourcing as an effective means of anticipating algorithmic impact which complements other approaches towards assessing algorithms in society by leveraging participatory foresight and cognitive diversity. 
\end{abstract}

%%
%% The code below is generated by the tool at http://dl.acm.org/ccs.cfm.
%% Please copy and paste the code instead of the example below.
%%

%\ccsdesc[500]{Concept~Concept}
%\ccsdesc[300]{Concept~Concept}
%\ccsdesc{Concept~Concept}
%\ccsdesc[100]{Concept~Concept}

%%
%% Keywords. The author(s) should pick words that accurately describe
%% the work being presented. Separate the keywords with commas.

%%
%% This command processes the author and affiliation and title
%% information and builds the first part of the formatted document.
\maketitle

% Intro

\vspace{-1mm}
\section{Introduction}

%RQ: How can we anticipate potential areas of impact of different Algorithmic Decision Making tools (ADMs) by employing crowdsourcing to leverage participatory ethics?

Algorithms have become commonplace in policy making, media, and even modern vernacular and discourse. There is increasing public interest in understanding how algorithmic decision making tools (ADMs) will affect individuals' lives and impact society, in addition to how they work \cite{mulligan2019procurement, Samek2019, GunningScienceRobotics2019}. Though designed typically for a particular purpose, these ADMs can have impact far beyond the initial considerations of the creators. Algorithms can be harmful--they have ethical weight, and therefore it is incumbent on designers and developers to anticipate, avert, and mitigate as many negative societal impacts as possible. Yet algorithm designers may not be aware of everything that could be impacted by their innovations. They may not consider a wide enough array of impacts for a variety of reasons: impacts are hard to predict due to uncertainty \cite{nanayakkara_anticipatory_2020}, designers' positionality to the problem at hand or to society may be limiting, and the incentives around disclosing probable impacts (absent a legislative mandate) may not be aligned \cite{bolger2017use, bonaccorsi2020expert}. 

This research explores the question of how to help anticipate potential areas of societal impact for ADMs. In particular we explore how the cognitive diversity of a crowd could help shed light on the many other areas likely to be impacted beyond the direct intention of the algorithm designers. There have been many efforts to encourage researchers and other algorithm developers to consider potential impact—asking them to write broader impact statements \cite{nanayakkara_unpacking_2021}, algorithm auditing \cite{diakopoulos_aa_2015, Raji:2019ti}, institutional review boards \cite{grady2015institutional, bozeman_science_2006}--but all of these happen late in the process, typically \textit{after} development and deployment. Anticipatory approaches, including participatory methods, are promising for integrating ethical considerations earlier into innovation processes \cite{brey_ethics_nodate}, but have been underexplored with respect to their efficacy for understanding the impact of algorithms in society. In this paper we aim to address this gap in ethical evaluation. The ability to achieve this task would help inform relevant stakeholders to algorithm design as well as help inform those concerned with the implementation of the ADM like policy makers, journalists, and the end users who it affects. Additionally, an effective tool to anticipate impact areas would help inform artificial intelligence ethicists and encourage algorithm designers to consider impact from the outset rather than as an afterthought.

%In order to escape this inherent bias of the algorithm designers and primary stakeholders,

In this work we explore a method to employ participatory ethics to anticipate algorithmic impact via crowdsourcing. In particular, we examine the applicability of using Amazon Mechanical Turk raters to describe potential areas of societal impact of various ADMs gathered from an online database of ADMs in use by the U.S. government. We use a contextualized topic model to group the various areas of impact described by the crowdsourced workers in ratings explanations and associate these topics to the ADM. Our analysis reveals that the crowd discovered four different types of topics: valence, general societal domains, intangible impact types, and algorithmic concerns like privacy and bias. On average with each additional evaluator up to the five we examined, we uncovered more diverse topics, however the increased number of impacts identified by the crowd began to plateau as more raters were included.

We make three concrete contributions in this paper. Primarily this paper evaluates crowdsourcing as a means of employing participatory foresight in order to anticipate different algorithmic impacts. Through this methodology we uncover 4 main types of impact (valence, societal domain, impact type, and algorithmic concerns), describe each of these areas in detail, and discuss potential limitations and gaps in the approach in comparison to other taxonomies of impact. Secondarily we explore the effectiveness of participatory foresight for achieving cognitive diversity by evaluating the added insights from each additional crowd worker. Finally, we analyze the complexities of the interactions between the different impact areas in order to provide insight into what patterns crowdsourcing can uncover in this manner that may otherwise be unobserved through other traditional methods of anticipating algorithmic impact.

% Lit Review

\vspace{-2mm}
\section{Related Literature}

We first explore the literature on algorithmic impact and current methods to evaluate impact in scientific research. Then we delve into the related work on anticipatory ethics and participatory foresight that laid the groundwork for our proposed approach to evaluating impact using crowdsourcing.
\vspace{-1mm}
\subsection{Algorithmic Impact}
In tandem with the growth of datafication worldwide comes the unprecedented growth of the use of algorithms to operate on that data at scale \cite{boczkowski_abundance_2021}. Such algorithms are now acting to inform decisions across many aspects of society, such as governments on policy decisions \cite{engstrom2020government}, doctors and medical professionals on advances in medicine and care decisions \cite{benjamens2020state}, news organizations on which journalistic pieces to create and showcase \cite{Diakopoulos_2019_news}, and even individuals on daily entertainment choices such as what music or TV shows to consume \cite{anderson2020algorithmic, gomez_netflix_alg}. Though they promise to bring about positive growth in society, algorithms have ethical weight and impact, both positive and negative, spanning a wide array of domains. 

There are a variety of approaches through which researchers and other external bodies have attempted to assess the societal impact of their proposed ADMs. One method that has recently been explored in academic conferences such as NeurIPS is through peer review\footnote{https://blog.neurips.cc/2021/12/03/a-retrospective-on-the-neurips-2021-ethics-review-process/}. Many grant awarders (e.g., the US National Science Foundation) also ask for societal impact to be considered during the peer review process. Reception of this additional step has been controversial, and some research argues that this is not the most effective way to assess impact \cite{holbrook_peer_2011} and may sometimes do more harm than good due to excluding potentially beneficial research that does not meet these criteria \cite{Roberts2009_societaLbenefit}. There are no set standards for this broader impact criterion (BIC), and even when researchers agree that impact should be considered prior to funding and publishing research, there are many alternative to do so such as having institutions set specific goals that are politically legitimized (e.g., encouraging women's research in STEM) and then funding these goals \cite{bozeman_broad_2009}. 

%A major concern of these grant providers enforcing their own societal impact evaluations lies in the direct link between funding and evaluation of societal impact. Tying societal impact of algorithms to grant funding reaffirms that pieces of scientific research are inherently political objects \cite{frodeman_intellectual_2009}. However, researchers have always had to and will continue to need to consider impact in order to maintain positive relationships with governments and other funding organizations; the more scientists broadly discuss impact in their research, the more science will continue to be funded in general \cite{pielke_beyond_1998}.

One concern of this BIC type of governance of societal impact evaluation is that scientists tend to solely focus on the positive impact of their research, or simply suggest education and outreach of whatever concept they are proposing \cite{frodeman_sciences_2007}. Another is that such approaches have been critiqued for being too late in the design process \cite{bozeman_science_2006, pater2022no}. There need to be approaches that encourage scientists to consider the negative impacts alongside the positive \cite{hecht2018s} and do so earlier in the process. Internal algorithm auditing is one means of impact assessment that forces algorithm designers to consider potential negative impacts from inception to implementation of their algorithms \cite{raji_closing_2020}. Requiring scientists to go through an auditing process rather than simply arguing the potential benefit of their research could be a more fruitful process for surfacing ethical issues while there is still time to address them before publication.   

%There are also limitations to institutionalized ethical approaches. Institutional Review Boards (IRBs) act as a formalized last line of defense to evaluate scientific research in terms of its impact on human subjects, preventing potentially dangerous research from moving forward. 

Another approach is the use of impact assessment tools, though many researchers tend to view them as more of a legal checkbox than a resource to guide implementation and design \cite{cashmore_introduction_2009}. Within the set of approaches towards machine learning transparency, algorithmic impact assessments (AIAs) are becoming a dominant form of evaluating impact. These AIAs attempt to elucidate the potential harms of ADMs while also coming up with practical steps to hold them accountable \cite{metcalf2021algorithmic}. These AIAs closely resemble other impact assessment tools within other disciplines such as environmental \cite{esteves2012social}, human rights \cite{kemp2013human}, or fiscal impact assessments \cite{burchell1985new}. The use of these tools underlines that third parties are essential to the evaluation of impact--these may be behavioral scientists, design researchers, or the communities impacted by these algorithms. Similarly, Ethics and Society Review (ESR) is a recent process that mandates an ethical and societal review in order to be eligible for funding; instead of using a review board at the end as a formality this involves earlier ethical checks when design changes can still be implemented. Every researcher involved in the study was willing to keep submitting their projects via this manner; there is demand in research for useful ethical evaluations of projects \cite{bernstein_ethics_2021}.

Prior research has synthesized taxonomies of ethical concerns related to algorithms. A review in 2016 by Mittelstadt et al. and later expanded in 2020 by Tsamados et al. consolidated these main ethical concerns into the following groups: (a) epistemic concerns of inconclusive evidence, inscrutable evidence, and misguided evidence, (b) normative concerns of unfair outcomes and transformative effects, and (c) traceability concerns \cite{mittelstadt2016ethics, tsamados2021ethics}. The epistemic concerns leading to improper implementation disproportionately affect people of color--aptly described as ``The New Jim Code'' by Ruja Benjamin \cite{benjamin_race_nodate}; if algorithms are trained on data containing racist or otherwise discriminatory themes then the resulting outputs will too, but under the guise of being impartial. Unfair outcomes can take the form of data-driven discrimination \cite{zarsky2013transparent} that can negatively affect groups based on misguided predictions. Regarding transformative effects, the defaults of an algorithm have been shown to be extremely important; these default values can determine the implementation of an algorithm for decades so these need to be considered with care \cite{mckenzie_recommendations_nodate, haque2019exploring}. Finally, within traceability there is the issue regarding where to place the moral judgement; it is crucial to be able to know the \textit{people} who will be held accountable when an algorithmic decision results in negative societal impacts \cite{hidalgo_how_2021, JANSSEN2016371, diakopoulos_accountability_2016, kroll2021outlining}.

While the prior work discussed here demonstrates that there is both a need and a range of approaches to assess societal impact and ethical issues in algorithms, one understudied area is in systematic participatory approaches that leverage the input of diverse stakeholders to assess impacts, which is the focus of this work. 
\vspace{-7mm}
\subsection{Anticipatory Ethics and Participatory Foresight}

It is difficult to consider all potential impacts of a new technology. A level of uncertainty will always be present in how a technology might evolve from idea to deployment in conjunction with human intentions and social dynamics \cite{nanayakkara_anticipatory_2020}. As a response, anticipatory ethics seeks to examine plausible and possible impacts via various ethical analyses and methodological approaches \cite{brey_ethics_nodate, sarewitz2011}. Anticipatory governance has also been described as a means to guide science and technology development in a humane direction, which assists with maintaining a positive relationship between science and society \cite{guston_understanding_2014}. A recent illustrative example of anticipatory ethics was applied to the emerging technology of synthetic media and used scenario-based approaches to consider the implications of different applications of deepfakes in elections, helping to elucidate complex ethical issues and determine their relative impact, as well as suggesting paths towards remediation \cite{diakopoulos_anticipating_nodate,johnson2021}. 

%In another case, For instance, employing systematic means of evaluating the uncertainty of how algorithms will be used allows researchers to better understand the possible impacts of their work in addition to helping them understand how the algorithm will (or might) be used or misused \cite{nanayakkara_anticipatory_2020}. 

%  Incorporating ethical evaluation into machine learning models does not have to happen in isolation of improving model performance; Delphi is a state of the art model explicitly trained on evaluating ethical dilemmas, and it showcases that failure to consider ethical implications of models vastly decreases the applicability to interact with humans \cite{jiang_delphi_2021}. Using morality as a feature in natural language processing tasks has even been shown to vastly enhance model performance on sentiment analysis and stance classification \cite{rezapour_enhancing_2019}. Incorporating ethics into algorithms is a symbiotic step for both society and the algorithms themselves.

Various methodological approaches such as scenario building and trend analysis can be applied in anticipatory ethics, but in this work we focus specifically on a \textit{participatory} approach as a way to leverage cognitive diversity for assessing ethical impacts \cite{brey_ethics_nodate}. Research in many diverse domains has examined the ability of cognitive diversity to lead to more informed decision making and improved trust in both the decision making process and the outcomes themselves \cite{Abby2015Cognitive, Bradley2007Management, FAUCHEUX2001223}. Placing the burden on one individual or a small group to consider every possible ethical angle of an ADM is a difficult task, especially when there are many conflicting views on what constitutes ethical decisions and fairness within machine learning and algorithms \cite{corbett-davies_measure_2018, Thomas1996, saxena2019fairness, hidalgo_how_2021}. There is no reason to believe that scientists or any other group of individuals have authoritative knowledge of what the ``social good'' is \cite{bozeman_broad_2009}. Furthermore, cognitive biases are prone to distort experts' predictive capacity of different impacts of their technology. One successful mitigation technique is through diversity: combining these expert opinions with those from other fields and lay-people with little or no expertise \cite{bonaccorsi2020expert}. Participatory foresight can be helpful in achieving a diverse view of anticipated outcomes to overcome the narrow minded lens of those in power determining algorithmic usage or design. Including diverse stakeholders can provide a much more realistic and socially grounded image of reality and impact \cite{nikolova_rise_2014}. 

While there are various methodological options for implementing participatory foresight from engaging stakeholders in participatory design processes to hosting envisioning workshops \cite{guston_understanding_2014}, in this work we examine online crowdsourcing as a potentially low-cost way to implement participatory foresight. While crowdsourcing platforms, such as Amazon Mechanical Turk, reflect their own biases, they still offer a good deal of demographic diversity in the types of people who contribute there \cite{hargittai2020}. This work therefore examines the potential utility, efficiency, and viability of applying crowdsourcing as a participatory foresight approach to evaluate algorithmic impact.

%We propose a method within anticipatory ethics termed ``participatory foresight" in which stakeholder involvement is leveraged in order to obtain cognitive diversity as a solution to consider multiple different ethical impacts of algorithmic decision making tools in order for the algorithm designer and other stakeholders to be well informed about potential areas of impact. 

 %Instead, ethics are constantly evolving and including more people in the evaluation of potential ethical impact can only result in a more comprehensive view; frameworks such as social choice ethics have been examined as a possible means for evaluating artificial intelligence ethics motivated by the notion that people should democratically have a say in issues that affect them\cite{baum_social_2020}. [add in work about how involing stakeholders improves science's relationship with society and thus including stakeholders has the added benefit of improving trust]

%While there is a host of related work on exploring how anticipatory ethics and governance can help understand hard to predict scenarios and better prepare for the future, (a) the limited use of it within algorithmic decision making has been successful yet underexplored, and (b) there has been little examination overall leveraging crowdsourcing as a means to employ it.

% Data and Exploratory Analysis

\vspace{-2mm}
\section{Data and Preprocessing}

\subsection{Data}
\label{sec:data}
We leverage data produced in the context of a system called Algorithm Tips, a computational news discovery tool designed to help journalists quickly find leads while reporting on the impact of algorithms in society by identifying newsworthy leads about the use of algorithms in government  \cite{diakopoulos2021towards}. The system scrapes government documents weekly and those documents are filtered and enriched through several annotation processes including a crowdsourced rating protocol. The government algorithms that are eventually selected for inclusion in the system are manually screened to meet a minimal definition of algorithmic-decision making\footnote{An algorithm is defined by the system as ``a set of rules to which data can be input and from which a result, such as a score, a calculation, or a decision, is obtained. Algorithms can be either computational (e.g., computer software or a spreadsheet) or not computational (e.g., a weighted score card or flowchart that could be applied by a person).''}.

The ADMs selected for inclusion are then augmented with metadata including jurisdictional information, a manually written summary, and by using a crowdsourcing protocol which leverages Amazon Mechanical Turk to employ five independent reviewers to quantitatively rate and explain their rating across four categories: controversy, magnitude, societal impact, and surprise. For instance, for the societal impact dimension workers are prompted with the statement: ``The algorithm described has the potential to create negative impacts in society'' and then asked to rate it on a scale from ``1'' (completely agree) to ``5'' (completely disagree) while also being asked to explain and justify their rating. Although the original context of this data collection is meant to support journalistic decision-making, in this work we repurpose this data to explore the potential for crowdsourced evaluation of impact. The open-ended crowdsourced explanations for ratings in tandem with the numerical rating comprise the primary data used in this research.

Prior to data validation, out data consists of 15,300 ratings (3,825 for each category) collected periodically from June 2019 to Jan. 2022 for 765 ADMs (5 ratings per ADM document). While the original data collection method already incorporated a number of crowdsourcing techniques and best practices to enhance data quality \citep{Wu:2017uh, McDonnell:2016we}, we found it necessary to develop an additional filtering step to remove any crowd responses that were of ``low quality'' in which the crowd worker had not adequately explained their numerical rating. We next describe our data filtering process used to remove these low-quality ratings and provide a more robust dataset for analysis. 

%The first phase of this project is to clean these ratings with a clear definition of what constitutes a ``valid” and ``substantial” explanation to both be substantive enough for humans reading the reviews and to feed the NLP model. After the definition has been applied by two independent reviewers (co-authors of this paper) with an adequate inter-rater reliability Cohen’s Kappa, we can train a model to automatically classify ratings as substantial and request another reviewer until five substantial independent reviews have been executed. This will have the added benefit of cleaning the data-stream for the algorithm tips tool as well as my own. Depending on the robustness of this model in relation to different fields of reviews, this could feasibly be used for validation of many different reviews beyond the scope of this project with minimal adjustment to the model.
\vspace{-2mm}
\subsection{Data Filtering}
To filter out low-quality (i.e. invalid) rating explanations we trained a supervised model using the texts of rating explanations. To train this model and apply it at scale, we first established a training dataset by labelling a sample of rating explanations as valid or not. To do so we applied the below criteria for defining validity. The explanation:
\begin{itemize}
    \item has to answer the question of specifically ``why'' the evaluator chose the ranking they did.
    \item cannot be a variation of ``it depends'' without a specific elaboration for what would validate a more confident rating.
    \item cannot be a nonsensical/unintelligible statement.
\end{itemize}

We started with a set of 400 randomly sampled ratings (100 each for controversy, magnitude, societal impact, and surprise), and had two raters evaluate the validity of the explanations. To assess the reliability of the rating process we computed the inter-rater reliability between the raters and reached a Cohen’s Kappa of 0.63, with varying scores within the four categories ranging from 0.48 to 0.81. The raters then discussed the definition and any differences in ratings and evaluated a new randomly selected sample of 200 ratings (50 for each category). After the second round, the inter-rater reliability score was improved to an acceptable level (Cohen’s Kappa of 0.8, ranging from 0.71 to 0.85). One rater then evaluated an additional random sample of 1900 reviews to arrive at a dataset of 2,500 explanations in total (625 for each of the four dimensions).

We reserved 500 of these explanations (randomized but stratified equally among the four categories) for the test set, and then trained a model on the remaining 2,000 explanations using 5-fold cross validation. To minimize false positives we optimized the model on an F-beta score of 0.5 in order to emphasize precision. We experimented with a variety of models (Naive Bayes, Logistic Regression, Random Forest Classifiers, and Random Forest Linear Regression) with diverse feature spaces (unigrams, bigrams, length, various dictionary similarity metrics). We ultimately used features based on several dictionaries to reduce overfitting to the training set, and selected `Length', `Quant', `Numbers', `Humans', `Affect', `Posemo', `Negemo', `Cause', `Health', `Money', `Death', and `Eval' from the LIWC2007 framework dictionaries \cite{LIWCDict_Citation}. We experimented with building separate models for each of the four ratings, but that led to unnecessary complexity and overfitting.

The final model developed leveraged features including unigrams, bigrams, length, and maximum cosine similarity scores between pretrained GloVe word embeddings \cite{pennington2014glove} for each word in the rating explanation text and each word in each of the dictionaries listed above. This model had a F1 score of 0.869, F0.5 score of 0.839, Precision of 0.819, Recall of 0.925, and a test accuracy of 0.869. 

%The model performs fairly consistently across the four rating categories (Controversy, Magnitude, Societal Impact, and Surprise), with slightly higher than average false positives for surprise. When it comes to rating on the 1-5 scale, the model classifies more threes as invalid than any other number, but it is hard to say if that is due to the model itself or the inherent nature of people who pick the central rating category simply not providing an insightful justification as to why.

%Assuming the cost of a review is \$0.60, there are 10 algorithmic decision making write ups scraped per week, and we have five reviewers initially evaluating the ADM tool, we can calculate the additional cost of using this model. Since the total negative rate was 28\%, after running the model on three iterations of evaluators we will have on average 49.44 valid sets of explanations (after initially seeking 50), which will cost an additional \$13.20 (an additional 44\% cost from the initial investment of \$30). It is important to note that only 17\% of this cost is from false negatives--83\% is from true negatives that were originally just clouding the dataset. 

We applied the trained model to our entire dataset of 15,300 explanations and were left with 10,291 (67\% of the original set). These were spread out relatively uniformly across the five point rating scale with the exception of `3' having about half the valid explanations as the other 4 rating levels. Valid ratings were dispersed fairly uniformly across the four rating categories, with controversy, magnitude, societal impact, and surprise receiving 24\%, 28\%, 25\%, and 23\% of the valid explanations respectively.
 
 After preliminary analysis with this filtered dataset, we decided to scope our findings to the 2,560 societal impact reviews due to their specific relevance to our research question. We initially evaluated if the other categories (controversy, magnitude, and surprise) were able to uncover other impact areas and meaningful insights, but ultimately they resulted in low topic coherence. The subsequent analysis therefore uses the filtered subset of data corresponding to the societal impact ratings and explanations. With this filtering each of the 745 ADM documents in our original dataset had 3.43 ratings on average (SD = 1.21). 96\% of ADMs had at least 2 ratings, 77\% of ADMs had at least 3, 53\% had 4, and 20\% had 5 ratings.

% Methodology       

\vspace{-2mm}
\section{Methodology}
\label{sec:methods}

\subsection{Contextualized Topic Models}

In order to identify different impact areas uncovered by the crowdsourced reviewers of the ADMs, we utilized topic modeling, an unsupervised natural language processing approach to group bodies of text into coherent topics. After evaluating multiple types of models, we chose a contextualized topic model--specifically CombinedTM \cite{bianchi-etal-2021-pre}. Contextualized topic models excel at improving coherence of the topics, which was critical for our analysis of identifying well-articulated impact areas. Most traditional topic models operate using bag-of-words inputs which do not utilize the context of the syntactical and semantic nature of sentences. Topic models benefit greatly from this contextual information, so incorporating it in the inputs tends to result in greater coherence. CombinedTM achieves similar accuracy scores to other state-of-the-art models while substantially improving on topic coherence and computational efficiency. 

CombinedTM incorporates Sentence-BERT (SBERT), or ``Siamese Bidirectional Encoder Representations from Transformers'', embeddings to add this contextual information to the bag-of-words topic model. SBERT improves upon the otherwise state-of-the-art BERT embeddings by dramatically increasing computational efficiency for various NLP tasks, especially unsupervised learning tasks like topic modeling \cite{reimers2019sentence}. SBERT's sentence limit of 512 tokens did not affect our corpus, which had a maximum length of 112 tokens.

%CombinedTM also employs a neural inference network that maps documents to approximate posterior distributions in order to improve the efficiency of training topic models by not requiring an entirely new derivation of each time the model is changed. The model used is called Autoencoded Variational Inference for Topic Model (AVITM) \cite{srivastava2017autoencoding}, which achieves similar accuracy scores to other state of the art models while dramatically improving upon both topic coherence and computational efficiency.

\subsection{Choosing a Model}

The main difficulty with choosing an appropriate topic model for a corpus such as ours was determining the optimal number of topics. The two main criteria we balanced to choose the number of topics were topic coherence and topic diversity. We evaluated three topic coherence metrics (UMass coherence score\footnote{$C_{UMass}(w_i,w_j) = log[\frac{D(w_i,w_j)+\epsilon}{D(w_i)}]$ where $D(w_i,w_j)$ is the number of documents containing $w_i$ and $w_j$ and $D(w_i)$ is the number of documents containing $w_i$; all computed over the original corpus resulting in an intrinsic measure}\cite{stevens2012exploring}, UCI coherence score\footnote{$C_{UCI}(w_i,w_j) = log\frac{P(w_i,w_j)+\epsilon}{P(w_i)\cdot P(w_j)}$; a pointwise mutual information (PMI) between two words over an external corpus resulting in an extrinsic measure}\cite{stevens2012exploring}, and normalized pointwise mutual information score\footnote{$NPMI(w_i;w_j) = \frac{PMI(w_i;w_j)}{-log[p(w_i,w_j)]} = \frac{log[p(w_i)p(w_j)]}{log[p(w_i,w_j)]}-1$}\cite{lau2014machine}) and evaluated topics that scored high on all three scores. For topic diversity we used the Inverted Rank-Biased Overlap (RBO)\cite{terragni2021word} and overall topic diversity (defined as percent of the top 25 words in each topic that are unique to that topic). We examined 100 topic models, incrementing the number of topics to uncover from 1 to 100. Across these models, the number of topics that resulted in the highest balance of these metrics (all of the coherence and diversity scores ranking in the top ten of the 100 models evaluated) were 7, 38, and 53. After training separate models with these number of topics and analyzing the results manually, 53 had the most diverse, coherent, and intelligible topics so we moved forward with this model.

We then went through the 53 topics and manually mapped them to 21 larger ``umbrella'' topics by following the methodology of computational grounded theory \cite{nelson2020computational}. Through this process we first allowed the unsupervised topic model to uncover the 53 topics as specified in the previous step, which computational grounded theory would term as ``Pattern Detection using Unsupervised Methods''. For the second step, ``Pattern Refinement using Guided Deep Reading'', we evaluated both the top words by uniqueness and frequency within each of the 53 topics in tandem with reading multiple rating evaluations that scored highly within the topics, and then grouped them into 21 topics based on similarity of content\footnote{Based on this step one topic of the 53 was omitted from further analysis due to being comprised of non-substantive content that tended to capture ratings that were phrased with ``I think that'' or ``I believe that'', rather than an anything more substantive.}. Finally for the last step, ``Pattern Confirmation using Natural Language Processing'', we then calculated again the coherence scores and topic diversity on the newly grouped topics. The topic diversity score went up substantially (which was to be expected since there is an inverse relationship between number of topics and topic diversity). Coherence scores, though lower than for the original 53, were still in the upper range (i.e., top decile) of the original topics evaluated. Because the additional human-driven reading and refinement added an element of coherence to the topics that the algorithms did not achieve, the subsequent analysis in this paper is based on the 21 umbrella topics developed using this process. An additional benefit of having undertaken a close reading of many of the documents in the corpus is that we are also better able to illustrate our analysis with salient examples of the topics. 

\subsection{Mapping Topic Model to Document Level Topics}

One main output of topic models are parameters assigned to each input document representing the proportion of each document assigned to a given topic (referred to as $\theta$s in topic modeling literature, but for the remainder of this paper we refer to these as document topic proportions). In our case, this corresponds to parameters assigned for each crowdsourced societal impact review of each ADM for the original 53 topics. The document topic proportions will sum to one for each ratings explanation document. The average document topic proportion assigned to each document across all topics was 0.0183 (median = 0.0098, and mean standard deviation = 0.0317). 

In order to prepare our data to map to larger umbrella topics and then further back to the original ADMs, we converted the document topic proportion to a binary flag of 1 if the parameter exceeded 0.050 (one standard deviation above the mean) indicating a topic was ``present'' in a review and 0 otherwise. In the case that no document topic proportion exceeded 0.05 (1\% of cases), we mapped the maximum value to 1 to have at least 1 topic per review. This resulted in a sparse binary array mapping each document to each of the 53 topics. To aggregate this array to the 21 umbrella topics, each umbrella topic was assigned a flag of 1 if any of its constituent topics had received a 1 (i.e., a logical OR of the sub-topic flags). Each ADM document in our dataset was thereafter associated with a set of binary values indicating the presence of each of the 21 umbrella topics in any of the ratings explanations associated with that ADM. 

%We then manually mapped these to the 21 larger ``umbrella" topics discussed above in a binary fashion: if a review had two of the 53 topics present that mapped back to one of the larger topics it only received a 1 for the topic being ``present", not 2. We then mapped all of these reviews back to the original ADM document ID, and we analyze these 21 topics by either rating or ADM for the remainder of this paper.

% Results

\section{Results}

The goal of this paper is to examine how to anticipate potential impact areas of different ADMs by employing crowdsourcing to implement participatory ethics. In this section we first go through each of the 21 topics and group them into four main thematic ideas in order to understand what types of impact the crowdsourcers can uncover: (1) valence, (2) societal domains, (3) impact types, and (4) algorithm concerns. We then analyze how the cognitive diversity afforded through using multiple crowdsourcers for each ADM relates to the number and types of impacts uncovered. Finally we examine how our data can uncover complex connections and relationships between impacts by analyzing co-occurrence patterns. 

\subsection{Topics Discovered}

We begin by describing the different types of topics that the crowdsourcers identified while evaluating the potential societal impact of the ADMs. The 21 ``umbrella'' topics we mapped to the 53 uncovered by the model (discussed in the previous section) were comprised of four different types of topics: (1) general valence, (2) societal domains, (3) impact types, and (4) more abstract concerns regarding the algorithm. All of these topic areas are listed in Table \ref{tab:topic_names} along with the overall proportion of the ratings the topic comprises (since ratings will usually have more than one topic assigned, this sums to more than 100\%), and the average numeric score the raters gave the algorithm on the 1 to 5 scale described in Section \ref{sec:data}. 

It is important to note that these results should be read in the context of the non-representative corpus of algorithms that we analyzed from the Algorithm Tips database, as detailed in Section \ref{sec:data}. Prevalences thus do not generalize to reflect algorithm use in U.S. government in general. Another bias of the data worth noting is that the overall skew of the ratings (on the five point scale) was overwhelmingly positive: the mean rating was 3.5, with a median of 4.0, and 61\% of the corpus was rated at a 4 or a 5.

% Other stats (if more interesting): 
% 1 = 7%
% 2 = 24%
% 3 = 9%
% 4 = 34%
% 5 = 27%

We next describe each of the four thematic groupings. Within each section we describe each topic individually in alphabetical order and provide example ratings and ADMs that had high document topic proportions for the given topics (we sometimes refer to these documents as `scoring highly' in a given topic).

\begin{table}[t]
  \caption{The 21 uncovered topics from the crowdsourced workers grouped into the four thematic areas of ``Valence'', ``Societal Domains'', ``Impact Types'', and ``Algorithm Concerns''. The topics are sorted in alphabetical order within the sections, and the percent of the ratings that discussed this topic and the average score on a five point scale are included.}
  \label{tab:topic_names}
  \begin{tabular}{lcc}
    \toprule
    \textbf{Topic Label}&\textbf{\% of Ratings}&\textbf{Avg. Score}\\
    \midrule
    \multicolumn{3}{c}{\textit{Valence}}\\ % align: l,c,r
    \midrule
    \textbf{Negative} & 5.6\% & 3.01\\
    \textbf{Neutral} & 46.1\% & 3.61\\
    \textbf{Positive} & 44.2\% & 3.87\\
    \midrule
    \multicolumn{3}{c}{\textit{Societal Domains}}\\ % align: l,c,r
    \midrule
    \textbf{Children} & 0.9\% & 2.58\\
    \textbf{COVID} &  8.2\% & 3.81\\
    \textbf{Education} & 1.4\% & 3.30\\
    \textbf{Environment} & 42.8\% & 4.14\\
    \textbf{Fraud Prevention} & 5.3\% & 3.90\\
    \textbf{Healthcare (Non-COVID)} & 14.4\% & 3.04\\
    \textbf{Infrastructure} & 7.1\% & 3.64\\
    \textbf{Mental Health} & 1.1\% & 3.38\\
    \textbf{Public Safety} & 5.2\% & 3.65\\
    \midrule
    \multicolumn{3}{c}{\textit{Impact Types}}\\ % align: l,c,r
    \midrule
    \textbf{Decision Making} &  8.7\% & 3.61\\
    \textbf{Efficiency} & 25.6\% & 3.79\\
    \textbf{Financial Costs} & 21.9\% & 2.63\\
    \textbf{Large Scope} & 5.5\% & 3.71\\
    \textbf{Risk Assessment} & 7.8\% & 3.80\\
    \textbf{Sustainability} & 8.1\% & 3.90\\
    \midrule
    \multicolumn{3}{c}{\textit{Algorithm Concerns}}\\ % align: l,c,r
    \midrule
    \textbf{Bias} & 9.5\% & 2.59\\
    \textbf{Harmful Results} & 9.5\% & 2.59\\
    \textbf{Privacy} & 24.8\% & 3.14\\
  \bottomrule
\end{tabular}
\end{table}

\subsubsection{Valence}

Valence topics were those that reflected an evaluation of how good or bad the ADM was for society (as opposed to specific areas of impact). 69\% of the ratings and 95\% of ADMs were assigned at least one valence topic through the crowdsourced reviews. On average, each review had at least 1 valence topic, and each ADM had 3 reviewers bring up a valence topic. The \textit{neutral} topic was the most prevalent and had an average rating of 3.61 whereas \textit{positive}, the next highest proportion, had a slightly higher average score of 3.87, and \textit{negative} was substantially lower with a score of 3.01. Given the positive skew of the numeric ratings in our dataset we leverage valence topics to calibrate those scores, taking the 3.61 average from the \textit{neutral} topic as the ``true'' neutral point for assessing the numeric rating. 

\textit{Negative} - Since we found that the ratings corpus generally skewed positively, the \textit{negative} valence topic was the rarest of the three (only 5.6\% of ratings). These ratings were overtly negative and typically failed to consider any potential positive impact. An example rating with a high topic proportion for \textit{negative} valence described the ADM as ``The tool could be designed with biases. Whether these are done subliminally or maliciously, they have the potential to negatively impact people of a specific gender, race, religion, and/or status.''
 
\textit{Neutral} - Ratings that scored high in the \textit{neutral} valence topic generally described what the reviewers perceived to be as standard procedures that they were not surprised were being implemented. They generally disagreed that these ADMs would have a negative impact on society (via the average rating of 3.61). An example of a document that scored highly on this topic is a risk assessment process that supports a community in understanding its natural hazard risks. A rating that scored high in this topic described the ADM: ``It streamlines an already objective process.''

\textit{Positive} - A substantial number of the reviews in the \textit{positive} valence topic were phrased in a manner such that the positive benefits outweigh the negative risks. These reviews generally took the form of ``The negative impacts would be xyz... but I think it has more positive impacts overall''. The majority of these types of statements were coded as positive instead of negative and therefore the positive topic has a substantially higher proportion of the corpus than the negative topic. An example of an ADM that scored highly on this dimension was designed to identify potential contaminating vulnerabilities of drinking water sources, and a rating that scored highly on this mentioned ``I think that it will have a positive impact on society by hiring more people that might not have the same opportunities as others because they are economically disadvantaged.''

\subsubsection{Societal Domains}

The different societal domains uncovered by the topic model represent various areas and aspects of society that the proposed ADM could affect. 67\% of the ratings and 94\% of ADMs were assigned at least 1 societal domain through the crowdsourced reviews. The average rating had 0.93 societal domains identified, and the average ADM document had 3.18 topics identified among all reviewers. On average, the societal domains of algorithms that scored the most positive rating were \textit{environment} (4.14), \textit{fraud prevention} (3.90), and \textit{COVID} (3.81), while the most negative ratings on average were in \textit{children} (2.58) and \textit{healthcare (non-COVID)} (3.04).

\textit{Children} - Ratings that involved children as a potential group of impacted people were typically among the most negative evaluations of societal impact. Some of these ratings tended to stress the need for human interaction in the algorithm, or were worried about potential dangers for children if the algorithm made incorrect predictions or was implemented incorrectly. A rating that scored highly in this topic is: ``It could create or find false positives, which could lead to children who are otherwise not in danger being removed from their existing family environments.'' An ADM that had ratings score high in this topic had to do with automating a human resources evaluation that could result in putting employees on leave; evaluators believed that changes in employment of adults could have negative impacts on their families.

\textit{COVID} - Ratings discussing COVID-19 were separated from other healthcare topics due to their specific focus on this virus. One ADM that scored high in this topic was a self-assessment tool that determines whether county government employees are certified to enter work sites of programs that were reduced because of COVID-19, and a rating that scored high described an ADM as ``attempting to keep people safe and ensure minimal spread of Covid-19. I see that it's attempting to prepare people for going on on trips and taking precautions when they return''.

\textit{Education} - Ratings that discussed the impact area of education ranged from discussing students' standardized tests to evaluation of teachers and universities. An ADM that ranked highly on this was a diagnostic tool that evaluated students' home language literacy and math skills and then provided accommodating instruction for them. A rating that scored highly in this topic evaluated the ADM as follows: ``Depending on how the algorithm treats literacy across various ethnic groups, the system can be biased toward more traditional learning styles. My concern is that the bias, while unintentional, could lead to negative impacts within larger societal groups.''

\textit{Environment} - Ratings describing the environment were some of the most prevalent in our corpus (perhaps reflective of the ADMs in the corpus), and ranged from climate change at large to specific county hazardous waste levels. One ADM document that ranked highly in this topic was a dashboard visualizing fire and accident risks within a given community used by the National Fire Protection Association, and a rating that scored highly in this topic said ``nothing but good can come out of knowing the level of air pollution''.

\textit{Fraud Prevention} - Ratings discussing fraud prevention ranged from the most common financial definition to identifying more general deceptive tactics in various industries. An ADM document scoring high in this topic described a predictive analytics technology to identify and prevent fraud, waste, and abuse in the Medicare fee-for-service program, and a rating that scored high described an ADM as: ``I think everyone would be glad to hear that things were being done to prevent fraud''.

\textit{Healthcare (Non-COVID)} - Ratings discussing healthcare (not specifically COVID) ranged from discussing hospital employees to specific illnesses. One ADM document that scored high on this topic discussed the effectiveness of Maternal and Child Health Bureau funded projects addressing prenatal and postpartum care; a rating that scored highly was ``the study mentions the need to evaluate individual cases, taking into account - among other issues - the substance use which has varying levels of intricacies, demonstrated the ever important need to look at each case individually, rather than trying to find a predictive, one-size fits all identification solution''.

\textit{Infrastructure} - Ratings discussing infrastructure ranged from assessing traffic patterns to government funding allocated to community improvement projects. One ADM document scoring highly in this topic was a scoring mechanism to assess pedestrian comfort in certain traffic conditions to indicate areas where infrastructure improvements needed to be made, and a rating that scored highly evaluated an ADM as ``Helping to rate flood risks can allow customers to better assess their risk of property damage, positively''.

\textit{Mental Health} - Ratings discussing mental health typically discussed the potential impact of ADMs on individual's happiness and mental stability, often in relation to jobs or other health areas. One ADM document that scored highly in mental health was a scoring formula to rate occupations used by a state's Department of Workforce Development to affect a program's eligibility to receive funding through the Workforce Innovation and Opportunity Act. A rating that scored highly in mental health described the following: ``While the thought of depression and suicide aren't pleasant and calculating one's risk of MDD might disrupt one's mood and actually lead them into depression, generally it is better for the person themselves and their family members to know about the risk and monitor their own behavior''.

\textit{Public Safety} - Ratings concerning public safety evaluated how implementation and use of the ADM would affect the safety of individuals or communities. One ADM document that scored highly in this topic was an online tool that leverages a user's self-reported symptoms to determine the likelihood they have COVID-19. A rating that scored highly in this area detailed: ``I rated this as Generally Disagree because I feel this has little to no chance of having negative impacts in society. I feel that this will allow people to stay safe, keep information honest and open, and give people the power to understand their community and where it is safe to travel''.

\subsubsection{Impact Types}

The different impact types uncovered by the topic model represent various ways or dimensions through which impacts could manifest. 62\% of the ratings and 94\% of ADMs were assigned at least 1 impact type through the crowdsourced reviews. The average rating had 0.77 impact types identified, and the average ADM document had 2.66 identified among all reviewers. On average, the impact types of algorithms that scored the most positive were \textit{sustainability} (3.90) and \textit{risk assessment} (3.80), while the only one with a substantially negative skew was \textit{financial costs} (2.63).

\textit{Decision Making} - Though all ratings are evaluating algorithmic decision making systems, ones that scored high in this topic specifically discussed the process of and the tool's ability to make those decisions. One ADM document that scored high on this was a system used by the EPA providing a numerical evaluation of risks to human and environmental health posed by uncontrolled hazardous waste sites, and a rating that scored highly evaluated an ADM as follows: ``I think it is important there is a set of rules to determine where the greatest need is for payment to treat Medicare patients''.

\textit{Efficiency} - Ratings discussing efficiency often described processes as being streamlined or general improvements of non-novel systems. One ADM that scored high in this topic was a US Agency for International Development tool that enhanced operational efficiency in partners by projecting their funds, and a rating that scored high mentioned that the ADM they were evaluating ``simply offers quicker insight into the development of a variety of power sources used to generate energy in the United States''.

\textit{Financial Costs} - Ratings discussing financial costs were among the most negative of all ratings (average numerical rating of 2.63), and they typically discussed costs to businesses, individuals, or taxes. One ADM document that scored highly on this topic was a performance assessment of clinicians used to determine Medicare payment increases or reductions, and a rating that scored highly on this detailed the following: ``this may affect people negatively if they do not allocate enough money for taxes''. 

\textit{Large Scope} - Ratings scoring highly in this category generally described the magnitude of people that would be affected by the ADMs in a wide-reaching manner (as opposed to only affecting a small subset of individuals or people in a specific context). One ADM that scored highly was an algorithm identifying what data to include in the census collection, and a rating scoring highly in this topic evaluated the respective ADM as follows: ``The danger with these algorithms is that they can affect our shopping habits, eating habits, shift power centres, and unhappiness''.

\textit{Risk Assessment} - This topic concerned ADMs that evaluated levels of risk in any industry. One ADM document that scored high in this topic was a survey system for high-risk chemical facilities to determine if the facility presents a high level of security risk, and a rating that scored high in this topic stated ``because the data and input is standardized from specific users, there are many benefits to `personalize' risk assessments for these users. . .enabling understanding of natural hazard risks also enables natural hazard preparation beforehand to minimize trauma or damage later''.

\textit{Sustainability} - Ratings in this area typically described the algorithm's ability to improve the sustainability of whatever facility was implementing it, or more generally impact societal sustainability. One document scoring high in this topic was an excel-based tool that uses a freight company's self-reported data to assess its freight operations, calculate its fuel consumption and carbon footprints, and track its fuel-efficiency and emission reductions. A rating that scored high in this topic said ``the algorithm is seeking sustainability as a goal, which would help society and those impacted by PEACH [Preserving Environment and Community Heritage]. In order to society to exist, we must create sustainability''.

\subsubsection{Algorithm Concerns}

The more abstract topics comprising the algorithm concerns were the rarest of the four thematic groups of topics. 40\% of the ratings and 76\% of ADMs were assigned at least one algorithm concern through the crowdsourced reviews. The three types of concerns uncovered in the corpus were \textit{bias, harmful results, and privacy}.

\textit{Bias} - This topic represents concerns over the original way the model was trained leading to improperly biased predictions. An ADM that scored highly on this topic scored the value based care (efficiency and effectiveness) of medical professionals and hospitals to then determine whether participants were eligible to avoid downward payment adjustments. This document scored highly on the impact areas of healthcare and financial costs, but it also had a high score in this bias category. One rating specifically that scored highly in this topic detailed that the algorithm could ``manifest itself in algorithmic bias and dangerous feedback loops''.

\textit{Harmful Results} - This topic dealt with the potentially harmful aftermath from both intended and unintended results of the ADMs. An example of an ADM scoring high on this category is an algorithm determining areas that were at high risk of fire within a given community---the review in question was concerned that this algorithm would distract fire fighters and the community at large from fire prevention in areas that were not caught by the algorithm; they believed total reliance on the algorithm was dangerous. One rating regarding an environmental algorithm that scored highly in this topic described the societal effect as ``...if the estimates are off, soil and plants could potentially not get adequate amounts of evapotranspiration.''

\textit{Privacy} - Ratings concerning privacy discussed concerns over either the data collected for the algorithm or results of an algorithm potentially putting individuals' data at risk or expose them to unwanted parties. An example of a document scoring high on this topic described a tool that determines the risk level for ineffective performance and non-compliance of Adult Education and Family Literacy Act funded programs; some reviews were concerned with where the data of individual workers or programs would go afterwards. A rating that had a high topic proportion parameter for privacy described: ``Some people may argue that such sensitive data and release thereof needs to be assessed by human employees to ensure that privacy is not unnecessarily violated''.
\vspace{-2mm}
\subsection{Cognitive Diversity}

\begin{figure}[h]
  \centering
  \includegraphics[width=\linewidth]{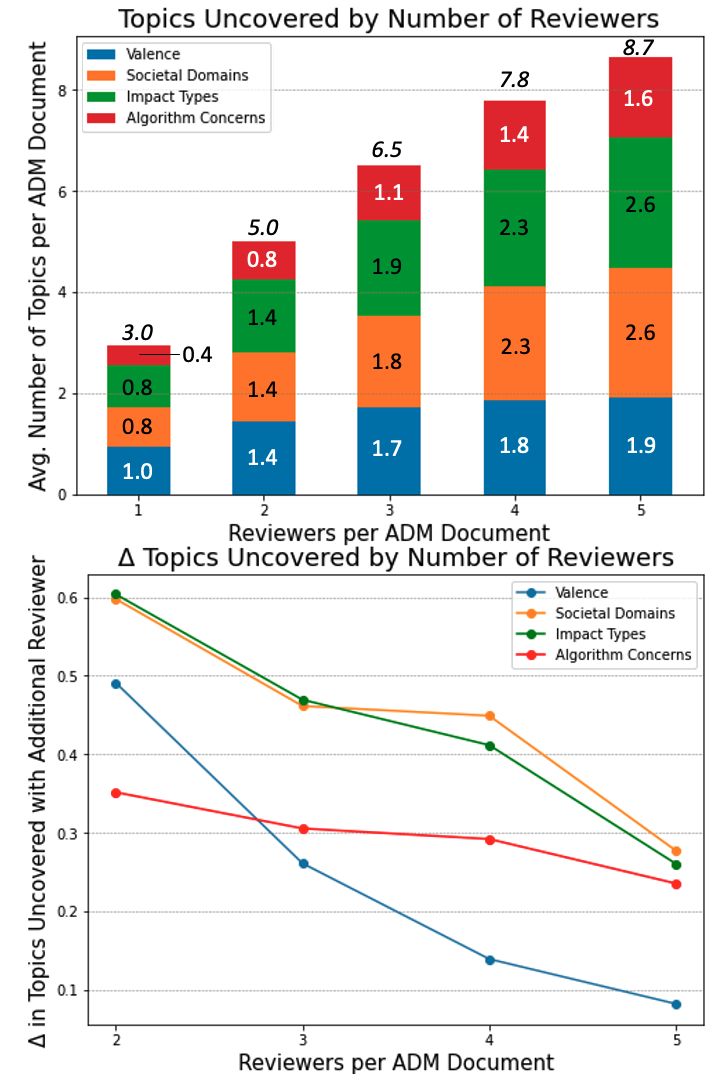}
  \caption{The top plot displays the average number of topics uncovered per ADM document by the number of crowdsourced reviewers that evaluated it on societal impact, separated by the four types of topics (Valence, Societal Domains, Impact Types, and Algorithm Concerns). The bottom plot shows the delta in the additional number of distinct topics uncovered with each additional reviewer.}
  \label{fig:cog_diversity_bar}
  \Description{Stacked bar plot of number of topics uncovered per reviewer included and line chart depicting the number increasing for each reviewer.}
\end{figure}

In this section we show how the cognitive diversity afforded through crowdsourcing ehances identification of algorithmic impact. We find that the use of crowdsourcing to evaluate the potential impact of the ADMs results in more potential impact areas uncovered, and shines some light on potential public sentiment and concerns regarding algorithm implementation. To understand how the different reviewers contribute to the topics uncovered in this analysis, we randomly order the number of reviews from a given ADM document (1 through 5) ten times, average the results, and analyze the number of impact areas uncovered with each additional review. As shown in Figure \ref{fig:cog_diversity_bar} in the top panel, on average one reviewer uncovers 3 topics, the addition of reviewer 2 almost doubles that to 5, then the growth starts to slightly decline with the inclusion of reviewers 3, 4, and 5 uncovering 6.5, 7.8, and 8.7 topics on average, respectively. The bottom panel of Figure \ref{fig:cog_diversity_bar} shows that the rate of discovery of all four types of impact areas declines as the reviewers increase, but at different rates. 

Valence topics decrease at the quickest rate and then plateau; after one reviewer you are likely to have one valence topic, the second reviewer on average uncovers 0.5 more and then it drops to 0.25 more or less with each additional reviewer. If you ask 3 reviewers you will on average get about 2 different valence topics. Societal domains and types decline at similar rates; they both decline from about 0.6 new areas discovered with the second reviewer to about 0.4-5 with the third and fourth, and 0.3 with the last. Algorithm concerns have a steady growth of about 0.3 new concerns per reviewer. These results suggest that five reviewers may be sufficient to survey in order to get a robust evaluation of different impact types and domains, though for all except the valence topics there could still be a trickle of additional impacts found with more reviewers. We conducted a validation study on a small subset of leads with more than 5 ratings each and found that with the exception of impact types, all of the other areas appeared to saturate at five reviewers.

One ADM document that estimated potential economic effects of climate change in the U.S. demonstrated this effect of more reviewers achieving cognitive diversity quite well. If we examine just one random ordering of the five reviews we uncover the following topics: Reviewer 1: \textit{environment}, \textit{positive}, \textit{financial costs}, the addition of reviewer 2 would describe \textit{bias} and \textit{neutral}, reviewer 3 would add \textit{efficiency}, reviewer 4 \textit{privacy}, and reviewer 5 \textit{decision making}. Though 3 reviewers all discussed the two topics of \textit{environment} and \textit{positive}, suggesting that those are perhaps dominant or consensus impact topics, the additional reviewers provided a much fuller picture by suggesting 6 additional impact topics that may be relevant. We next consider what we can further learn by examining overlaps and co-occurrences of topics, further underlining the utility provided by employing participatory foresight via crowdsourcing. 

% POTENTIAL EXTRA ANALYSIS: Quantify overlapping topics, e.g. "enviornment if discusses is always uncovered"

\subsection{Impacts in Relation to Each Other}

\begin{figure}[h]
  \centering
  \includegraphics[width=3.55in]{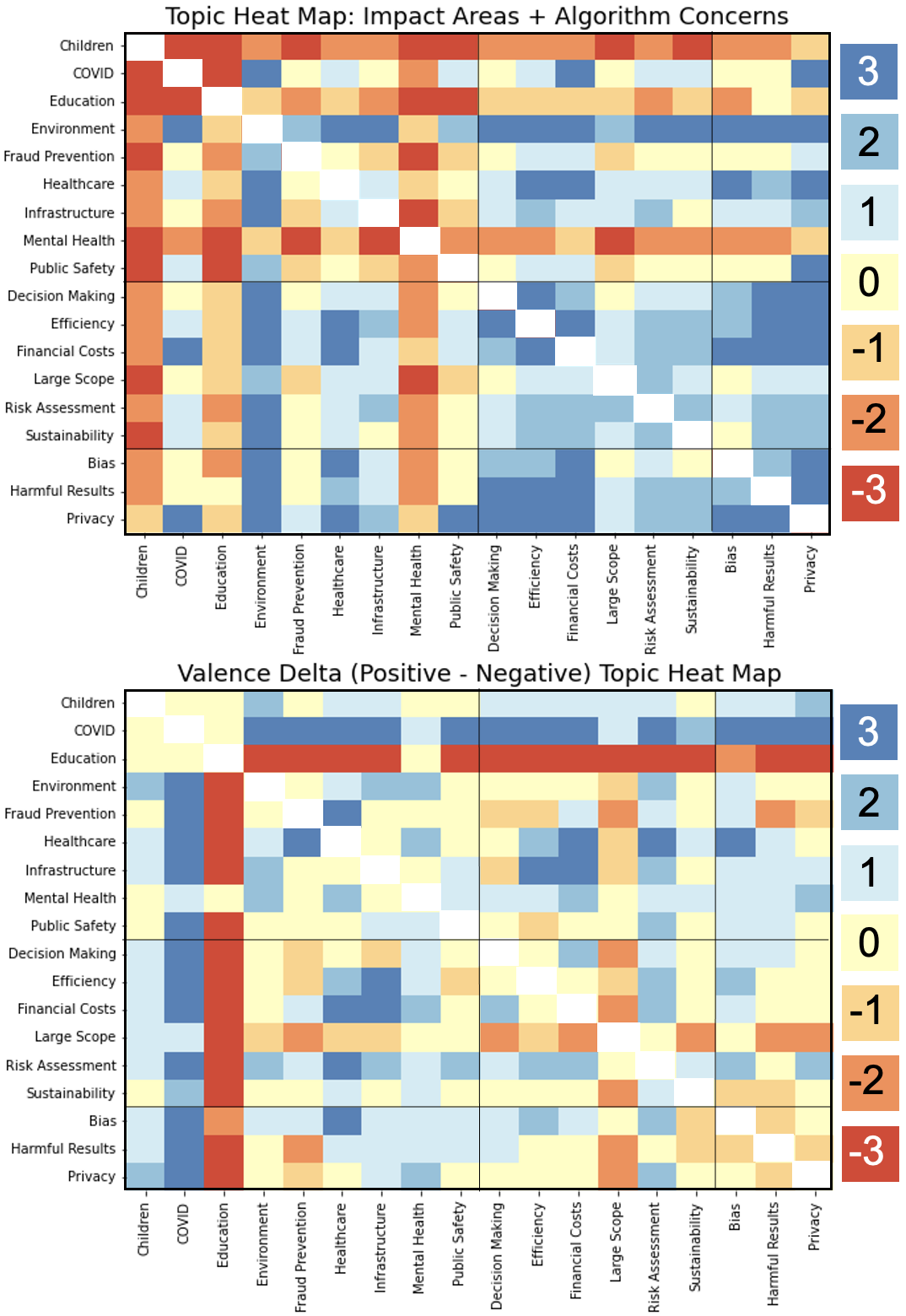}
  \caption{Two heat maps: (1) the relative co-occurrences of topics, and (2) the difference in relative co-occurrences within \textit{positive} and \textit{negative} valence documents (labeled as having at least one reviewer uncover a positive or negative topic, respectively) on a normalized scale.}
  \label{fig:heatmaps}
  \Description{Heat Map of Topics}
\end{figure}

In order to understand the different types of insight that crowdsourcing can provide to evaluate algorithmic impact, it is useful to explore how these topics tend to co-occur within reviews of ADM documents since there are notable trends across the impact areas that can be extended beyond any specific document. This gives further credence to the method of crowdsourcing to evaluate different interacting societal impacts because we can illuminate complexities in the connections between impact areas that may not be obvious when only evaluating a singular ADM document. We define co-occurrence as each time two topics were uncovered by reviewers in the same document. The top panel of Figure \ref{fig:heatmaps} shows a heat map of topic co-occurrences within ADM documents (on a standardized scale of the co-occurrences with the median mapped to 0 and the other bins above and below mapped to evenly spaced quantiles). The bottom panel of Figure \ref{fig:heatmaps} shows a heat map of the difference in these relative co-occurrence scores between documents with at least one \textit{positive} or \textit{negative} valence topic. This helps show the relative interaction of topics with respect to the valence dimension.

Unveiling these types of topic co-occurrences to the algorithmic designer could unveil areas of impact that the designer had not otherwise considered. For example, an algorithm designed to determine if certain factories were at elevated risk for inspection scored highly on \textit{risk assessment} and \textit{public safety} which were presumably top of mind for the algorithm designer. However, this ADM also scored highly on \textit{large scope}, \textit{mental health}, and \textit{healthcare (non-COVID)} because some of the reviewers brought up that if this causes factories to shut down unnecessarily then workers could be out of jobs and suffer poor mental health and risk losing insurance. 

The first heat map in the top panel of Figure \ref{fig:heatmaps} is skewed by the overall proportion of a topic in the corpus (for instance, \textit{environment} was in 43\% of the corpus and thus co-occurs with most other topics whereas \textit{children} was only 1\% of the corpus and co-occurs relatively infrequently with everything), but interesting trends appear when we compare these relative co-occurrences within the valence categories of \textit{positive} and \textit{negative}.

The top panel of Figure \ref{fig:heatmaps} illustrates that the topic types with the greatest rate of co-occurrence are impact types (e.g., \textit{decision making}, \textit{efficiency}) and algorithm concerns (e.g., \textit{privacy} and \textit{harmful effects}, and impact types within themselves. They occurred together at a higher rate than societal domains did with algorithm concerns, though there was no particular valence trend within these co-occurrences. Societal domains had a more infrequent rate of co-occurrence with themselves than any other type.

As shown in the bottom panel of Figure \ref{fig:heatmaps}, algorithm concerns had an elevated number of interactions with impact types within positive documents, and a relatively stronger positive valence when interacting with societal domains. Individuals perceived most topics' interactions with \textit{education} in a more negative valence than they did with any of the other impact areas, whereas there was a stronger positive valence when they occurred with \textit{COVID} or \textit{healthcare}.

% \textit{Education} and \textit{large scope} were two areas that had a stronger correlation with \textit{negative} documents in general; reviewers typically thought that algorithms in these areas had a strong potential for negative impact on society. Notably, among the highest negative pairings were these two impact areas with the algorithmic concerns of \textit{privacy} and \textit{harmful results} as well as when they co-occured, indicating that individuals viewed these as having an even greater negative societal impact when they were both potential outcomes of the algorithm. The topics that had a much higher relative co-occurrence with \textit{positive} documents were \textit{COVID} and \textit{healthcare} with the exceptions of interaction with \textit{education} and \textit{decision making}; reviewers typically evaluated these algorithms favorably (which falls in line with the average score seen in Table \ref{tab:topic_names}.

Some topics like \textit{fraud prevention} and \textit{financial costs} had a notable dispersion across \textit{positive} and \textit{negative} valence depending on what topic was also discussed. For instance, when \textit{fraud prevention} was discussed in tandem with \textit{education} or \textit{large scope}, it was more likely the document was discussed with a \textit{negative} valence. However, when discussed in light of \textit{COVID} or \textit{healthcare}, there was a more dominant tendency to appear on documents evaluated with a \textit{positive} valence. This indicates that the intersection of topics has an important effect on societal evaluation of algorithms.

% Discussion

\vspace{-2mm}
\section{Discussion}

% Big Q: What does this all mean for the participatory ethics approach? Should we be making changes? Are there clear limitations? 

% Step back, go over results, what do they mean, how do they relate back to research question, relate back to lit review, did we solve the gap, what future work needs to be done in this space based on what we've done. are there future directions of research based on what we've done?

% What does this all mean?

Our findings provide evidence that crowdsourcing can be an effective means of employing participatory foresight to anticipate different impacts of algorithmic decision making tools. Crowd workers were able to identify a variety of aspects across basic valence, impacted societal domains, impact types like efficiency or cost, and specific algorithm concerns like bias or privacy. The range of impacts identified increased with more crowd workers. Based on our analysis we next elaborate limitations with our specific approach, contrast our findings to an ethics taxonomy, and elaborate how our method helped harness cognitive diversity.  

We first acknowledge the limitations of our specific data source. Although we were limited by the way in which the Algorithm Tips database was constructed, we recommend that future research leveraging this approach to examine variations in the open-ended question wording for data collection. For instance, more specific survey questions might probe at specific impacts like bias or privacy for more detailed concerns (e.g. biased against particular stakeholders). As noted, the corpus analyzed is a non-representative sample and so has its own biases due to the way it was collected. Due to positive valence skew, it would be illuminating to ask individuals to explicitly list the positive and negative impacts rather than just the degree to which they believe the ADM would create negative impacts. This could result in more information, more impact areas, and a wider array of valence evaluations. Additionally, all of the ADM documents we evaluate are already in stages of deployment; although a motivation for the work, our analysis did not directly address the early stage of the design of these tools. Additional analysis would be needed to see if crowdsourcing could have as useful of insight at earlier stages of algorithmic design in order to have influence on the design of the algorithm at the earliest stage possible while the technology could still be steered and adapted \cite{brey_ethics_nodate}.

Looking at the different areas of ethical concern related to algorithms established first in 2016 by Mittelstadt et al. \cite{mittelstadt2016ethics, tsamados_ethics_2021}, our methods uncover some but not all of established ethical concerns of algorithmic impact. First, we were able to uncover specifically ``misguided evidence'' (\textit{biased data}), and less specifically but still effectually ``inconclusive evidence'' and ``unfair outcomes'' (\textit{harmful results}). We were unable to concretely map ``inscrutable evidence'', ``transformative effects'', and ``traceability''. ``Inscrutable evidence'', or lack of understanding of how the algorithm generates predictions and outcomes, is generally outside of the frame of reference for non-expert crowdsourced workers to evaluate ADMs for which they are only shown a brief description. Transformative effects could theoretically be achieved through this approach if the right question is asked, but we found that the crowdsourced workers typically thought on a smaller scale than transformative societal change--if a different question is tailored to uncovering this or a population with more technical expertise is queried, this could potentially be achieved. Similarly, ``traceability''  was beyond the scope of the task; we did not ask the crowdsourcers who they would hold accountable if the algorithm resulted in harmful effects, but a different set of questions might probe this factor as well.

We find that crowdsourcing is an effective means to achieve cognitive diversity of potential impact, because on average one reviewer only describes 2-3 areas of impact whereas if you ask five reviewers you uncover 4 times that on average. Without additional qualitative research we acknowledge that it is difficult to discern the true quality of this cognitive diversity (potentially just cognitive volume), but end-user stakeholders such as researchers or algorithm designers could evaluate the quality and act accordingly. Future work could examine after how many reviews the phenomenon of diminishing marginal returns truly sets in. Our method of analyzing 5 reviews per ADM indicated saturation for the valence dimension but the other dimensions, particularly impact types, might still benefit from obtaining more reviews. This research also underlines the fact that it is feasible to add a more wholly formed understanding of potential societal impact of different ADMs through the use of querying non-expert individuals. We did not explore the quality and quantity of findings uncovered by querying experts in other fields, like journalists, technical experts, social scientists, design researchers, or ethnographers in the field \cite{bernstein_ethics_2021}, but this could be a useful point of future research. It could also be useful to target different populations (e.g., minority groups or external experts) to ask what they think about these systems. By altering the sample of documents and evaluators future work can understand better the different impacts on society at large.

%We also laid the foundation for researchers to consider potential impact areas without taking the extra step of surveying crowdsourced workers to employ participatory foresight. It is important for algorithm designers to be aware of these trends to consider potential public perception when they are in the early stages of designing their ADMs. For instance if someone is designing an algorithm involving education, they need to be aware that the public is going to be wary of the algorithm and try to preemptively address concerns that may arise in the design of the algorithm rather than after the fact.

% Conclusion

\vspace{-3mm}
\section{Conclusion}

In this paper we evaluated crowdsourcing as a means for employing participatory foresight to anticipate different impact areas of algorithmic decision making tools. Our findings indicate that crowdsourcing can be an effective way to leverage cognitive diversity by asking multiple individuals to evaluate the potential societal impact of different ADMs. Through computational grounded theory we were able to establish 21 different topics across valence, societal domains, impact types, and algorithm concerns. We demonstrate that with the addition of each reviewer, more impact areas come to light and suggest that this diversity of thought could enable algorithm designers to consider a wider array of potential impact. We evaluate trends among the different areas when they co-occur within documents and in doing so give further credence to the method of crowdsourcing to evaluate various interacting societal impacts, shedding light on how the complexities of the impact areas interact within different algorithmic decision making documents. We contrast our findings with what previous literature has identified as concrete ethical considerations of ADMs in order to elucidate the potential and limitations of using crowdsourcing in this manner.  Ultimately we demonstrate that crowdsourcing can be an effective means of employing participatory foresight in order to anticipate potential algorithmic impact, while also pointing out the variety of future directions opened up by this line of research. 

% ACKNOWLEDGEMENTS

\begin{acks}
This work is supported in part by the National Science Foundation via award IIS-1845460. We would also like to thank Jack Bandy and the reviewers for their helpful comments.
\end{acks}

%%%%%%%%%%%%%%%%
% References
%%%%%%%%%%%%%%%%

\bibliographystyle{ACM-Reference-Format}
\bibliography{references}

%% APPENDIX

%\input{appendix}

\end{document}